\documentclass{ijuc}

\usepackage{amsmath,amsfonts,amssymb} 
\usepackage{graphicx} 
\usepackage{physics}
\usepackage[utf8]{inputenc}
\usepackage{float}
\usepackage{xcolor}
\usepackage{xcolor}

\usepackage{fancyhdr}

\usepackage{geometry}
\begin{document}
\thispagestyle{plain}

\title{Cascading of Nanomechanical Resonator Logic}

\author{X. Jin\inst{1}
\and C. G. Baker\inst{1}
\and E. Romero\inst{1}
\and N. P. Mauranyapin\inst{1}
\and T. M. F. Hirsch\inst{1}
\and W. P. Bowen\inst{1} \email{w.bowen@uq.edu.au}
\and G. I. Harris \inst{1} 
}

\institute{School of Mathematics and Physics, The University of Queensland, QLD 4072, Australia}

\def\received{Received 28 January 2022}

\maketitle

\begin{abstract}
Nanomechanical systems have been proposed as an alternative computing platform for high radiation environments, where semiconductor electronics traditionally fail, as well as to allow improved gate densities and energy consumption. While there have been numerous demonstrations of individual nanomechanical logic gates leveraging the Duffing nonlinearity, the development of useful nanomechanical logic circuits depends strongly on the ability to cascade multiple logic gates. Here we show theoretically that cascading nanomechanical logic gates, where the output of one gate is fed into the input of another, is a complex problem due to the transient dynamics of the collective system. 
These transient behaviours can lead to undesired bit flips, which precludes cascading altogether. We then show that this issue can be circumvented by carefully initialising the system prior to computation. We illustrate these salient features through the modelled dynamics of two cascaded nanomechanical NAND gates. 
\end{abstract}
\parskip=7.5pt 
\newpage
\section{Introduction} 
Over the last 50 years, semiconductor electronics have evolved to become the basis of nearly all modern technologies. This evolution was primarily led by advances in fabrication techniques, enabling a reduction in transistor size of three orders of magnitude, and commensurate improvements in performance, cost, and integration \cite{Liu_CMOS_15, Pott_10}. However, for devices less than $100\ \mathrm{nm}$ in size, non-trivial physical interactions begin to limit operation and make further scaling difficult. For example, transistors of this size exhibit increased current leakage from quantum tunneling and thermal excitation which increases power consumption \cite{Pott_10, Frank_02}. Indeed, the energy dissipated per logic operation in present-day silicon-based digital circuits is a thousand times larger than the fundamental thermodynamic limit, known as the Landauer limit \cite{Landauer_IBM_1961,Pop_NatRes_2010}. Furthermore, miniaturized transistors are also extremely susceptible to degradation and errors in harsh environments \cite{Velazco_19}. Semiconductor electronics will undoubtedly remain a core component of future technologies, however, specialized applications requiring ultra-low power consumption or operation in extreme environments may require different computing platforms.

The susceptibility of semiconductor electronics to failure in high radiation environments is well documented \cite{Velazco_19}. When ionising radiation passes through electronic circuits, atoms are ionised and electron-hole pairs are generated. This can cause charge accumulation and mechanical damage, resulting in the gradual degradation of performance over time and eventually lead to catastrophic failures \cite{Kasprzak_15}. For instance, a breakdown in satellite communications is anticipated in the event of an intense solar flare, during which high levels of ionising radiation are emitted from the sun. It is estimated that such an event could completely destroy about $ 10 \%$ of all orbiting satellites, equating to approximately 70 billion dollars of damage \cite{Lingam_17, Odenwald_AdvSpaceRes_2006}. Furthermore, this problem is not unique to satellites, other technologies such as radiation therapy facilities and nuclear disaster clean-up robots are also exposed to extreme doses of radiation during operation \cite{houssay_PHD_00, Minwoong_MDPI_2020}

To improve the performance and reliability of computers in high radiation environments, many are now considering alternative computing architectures. 
%
One promising approach is to encode logical states into the motion of micromechanical elements, which are typically more immune to radiation exposure than semiconductor electronics \cite{Arutt_16}. In this context, there has been substantial work towards replacing transistors with microelectromechanical systems (MEMS), whereby the electrostatic actuation is engineered to induce displacements of suspended elements to enable computation \cite{Pott_10, Perrin_IEEE_21, Lee_Science_2010, Loh_NatNano_2012, Pillonnet_Springer_2017}. 
In parallel to this work, there is a sizable effort to develop purely mechanical computational systems which rely on mechanical interconnects and nonlinear mechanical oscillators to act as logic gates \cite{Roukes_04, Hatanaka_NatNano_2014, Malishava_PRL_2015, Feng_NatComm_2014}. 
Indeed, these purely mechanical systems are the foundation for the work presented here. 
Importantly, with the recent advances in nanofabrication, nano-sized mechanical elements can now be produced at volume and with unprecedented accuracy using CMOS compatible techniques \cite{Wolf_15}. 

For effective and scalable computation, nanomechanical computing architectures must have the following five essential characteristics \cite{Behin-Aein_10}:
\begin{enumerate}
\item Nonlinear behaviour: this enables digitization via distinct steady states (i.e. bistability).
\item Power amplification: the energy to switch logic states comes from a separate pump.
\item Cascadability: the output of one gate can feed into another.
\item Feedback prevention: the output of one gate does not affect its input.
\item Complete set of Boolean operations: the gate can perform logic operations required for universal computing (i.e. NAND or NOR).
\end{enumerate}

Several studies have experimentally demonstrated individual nanomechanical logic gates \cite{Song_19, Wenzler_14, Yamaguchi_11, Yao_14}. However, despite these initial successes, these designs have not been able to fulfil all five criteria. For these systems, various technical limitations prevented cascading of multiple nanomechanical gates into complex circuitry~\cite{Wenzler_14}. For example, \emph{Ilyas et. al.} demonstrated two nanomechanical gates that were cascaded by capacitively measuring the motion of one oscillator then using the current to electrically actuate the next oscillator~\cite{Ilyas_18}. This strategy works well for a small number of nanomechanical gates, but doesn't allow efficient scaling to larger complex circuits, mainly due to the large losses associated with mechanical-to-electrical conversion at each gate. A far more energy efficient approach to cascading is direct mechanical coupling from gate-to-gate. In this context, a new platform has recently been proposed that enables nanomechanical oscillators to be directly connected via single mode mechanical waveguides~\cite{Mauranyapin_21}. Motivated by this platform, here, we aim to theoretically investigate the cascading of mechanical logic gates via direct mechanical connection. 

Previous models of cascaded logic have generally only considered the steady-state of the system. Here, we examine the dynamical response. Surprisingly, we find that in common scenarios where the steady-state solution would indicate a success, the computation actually fails due to unwanted transient dynamics. This leads to the question of whether the temporal dynamics of nanomechanical computing architectures inherently prevent computation. We find that by carefully initialising the system prior to computation, these detrimental transients are mitigated and computation is restored.

\section{Duffing Oscillator}

The Duffing oscillator is a common choice of nonlinear oscillator to perform mechanical computation \cite{Song_19,Yamaguchi_11, Yao_14, Tadokoro_21}. This is because the behaviour of Duffing oscillators is relatively well-understood, and commonly found within many physical systems \cite{Ivana_11}. Mathematically, the equation of motion of a Duffing oscillator is given by the Duffing equation \cite{Schmid}:
\begin{align} \label{Duffing}
\frac{d^2 x}{dt^2} + \gamma \frac{dx}{dt} + \omega_0^2 x  + \eta x^3 &= \frac{F}{m}\sin(\omega t),
\end{align}
where $x$ is the oscillator displacement, $t$ is time,  $m$ is the mass of the oscillator,  $\gamma$ is the damping rate,  $\omega_0$ is the natural resonance frequency and  $\eta$ represents the strength of the nonlinearity. The oscillator is driven by a sinusoidal force of magnitude $F$ and frequency $\omega$. When $\eta =0$, equation \eqref{Duffing} reduces to the equation of motion of a simple harmonic oscillator. When $\eta > 0$, the oscillator is said to exhibit spring hardening, as it experiences a stronger restoring force with increasing displacement. To study the dynamics of Duffing oscillators, it is often convenient to write equation \eqref{Duffing} in its dimensionless form, given by \cite{Schmid}:
\begin{align} \label{dimensionless_Duffing}
\frac{d^2 \tilde{x}}{d \tau^2}+ Q^{-1}\frac{d \tilde{x}}{d\tau} + \tilde{x} +  \tilde{x}^3 = \tilde{F} \sin( \tilde{\Omega} \tau)
\end{align}
where $\tilde{x} = \sqrt{\frac{\eta}{\omega_0^2}}x$ is the dimensionless amplitude of motion, $Q = \frac{\omega_0}{\gamma}$ is the mechanical quality factor, $\tilde{F} =  F \sqrt{\frac{\eta}{m^2 \omega_0^6}}$ is the dimensionless drive strength at relative frequency $ \tilde{\Omega} = \frac{\omega}{\omega_0}$ and time $\tau = \omega_0 t$. This equation of motion exhibits a non-trivial driven response, but solutions can be obtained either analytically, in the limit of weak nonlinearity and high Q-factor at steady-state, or numerically \cite{Schmid}.

\begin{figure}[!ht]
\includegraphics[width= 0.98 \linewidth]{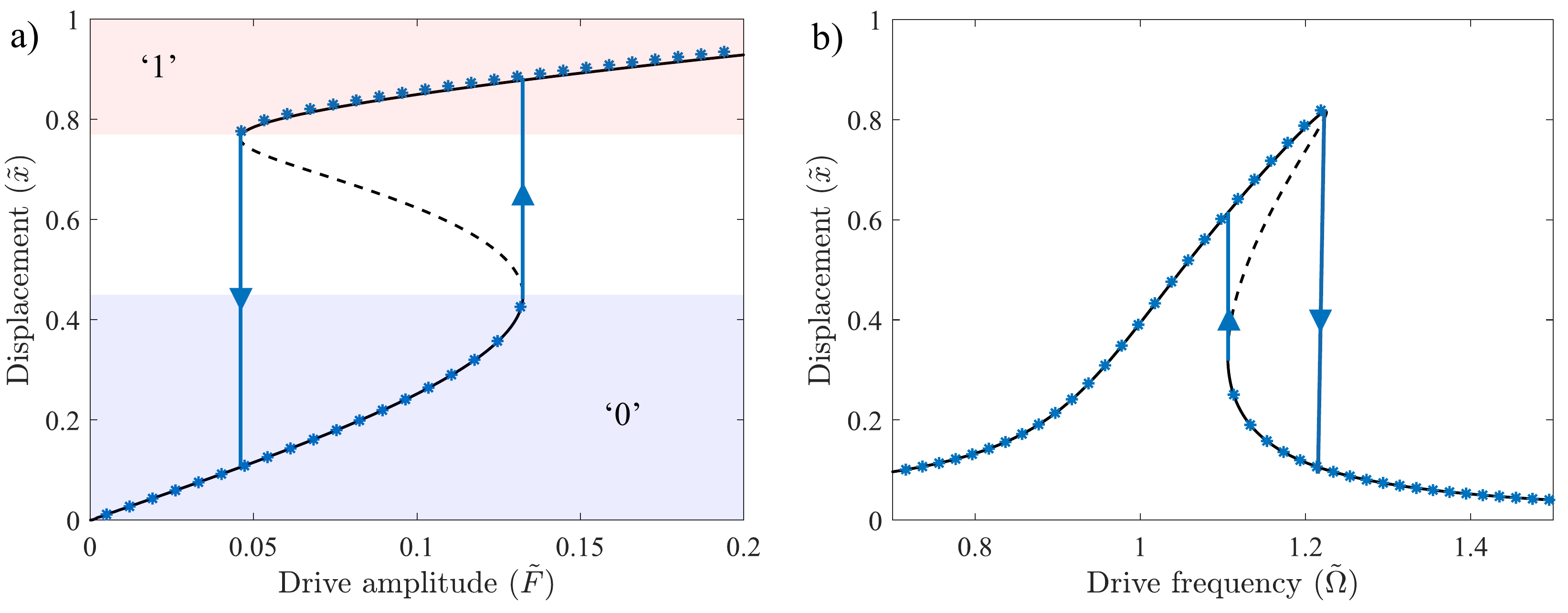}
     \caption{a) Drive response of a Duffing oscillator. The parameter values used to construct this response are $Q = 20$ and $ \tilde{\Omega} = 1.2$.  b) Frequency response of a Duffing oscillator. The same parameter values are used to construct this response, except with a fixed drive, $\tilde{F} = 0.05$, and a varying drive frequency. In both figures, the black curve represents the analytical solution and the blue data points represent the simulated results. The jump-up and jump-down dynamics are respectively represented by the up and down arrows. }\label{jumps}
\end{figure}

The characteristic driven response of a Duffing oscillator is multi-valued, with the equation of motion having two stable solutions and one unstable solution. This is shown in Figure~\ref{jumps}a), where the response of a Duffing oscillator exhibits hysteresis as the drive amplitude is increased, then decreased \cite{Brennan_07}. The sharp transition between the two stable solutions is called a \emph{jump} and the drive at which this occurs is called the threshold drive, denoted as $\tilde{F}_{c,up}$ and $\tilde{F}_{c,down}$ for jump up and jump down, respectively.  This multi-valued response allows displacement to represent the logical bits of a nanomechanical gate; high displacement outputs can be defined as binary `1' signals, and low displacement outputs can be defined as binary `0' signals, as labelled in Figure~\ref{jumps}a). The jump phenomenon can also be observed in the frequency domain, as shown in Figure~\ref{jumps}b), which exhibits hysteretic behaviour as the drive frequency is increased, then decreased.

The analytic results in Figure~\ref{jumps} are obtained by the method of harmonic balance~\cite{Brennan_07}. However, this method only yields accurate steady-state solutions for oscillators with a high $Q$ factor ($Q \gg 1$) that are driven near resonance ($\tilde{\Omega} \approx 1$) \cite{Schmid, Brennan_07, Wawrzynski_21}. Such analytical solutions do not provide insight into the temporal dynamics of Duffing systems, which are necessary to fully appreciate the behaviour of cascaded systems. Due to this, we have henceforth chosen to study the dynamics of the Duffing oscillator using numerical methods.

\section{Single Mechanical Logic Gate}

\begin{figure}[hbt!]
\includegraphics[width = 0.88 \linewidth]{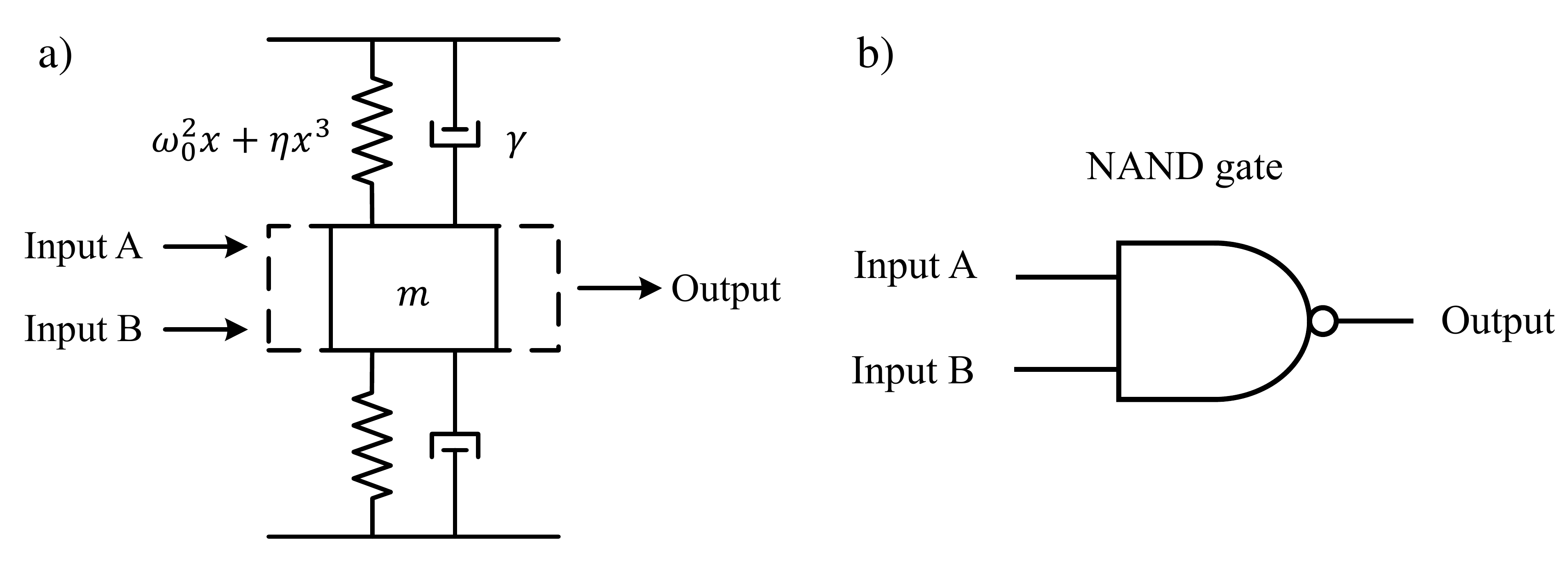}
\caption{a) Schematic of a single mechanical logic gate. The inputs/output of the gate are labeled and represented by arrows. b) Conventional symbol of a NAND gate. Note: both the schematic and the symbol only account for the two logical inputs, $A$ and $B$, but not the pump. }
\label{NAND diagram}
\end{figure}

In addition to assigning mechanical amplitudes to logical states, a single Duffing oscillator can also be configured to perform logic operations. In this study, we consider a Duffing oscillator driven by three separate drives, respectively denoted as $A$, $B$ and $P$. These drives will combine together, with some predetermined phase, into the forcing term $F$ of Eq.~\ref{Duffing} (i.e. $|F|=|F_A + F_B + F_P|$). 
Here, $A$ and $B$ are the inputs to the logic gate and can be assigned either a binary `0' or `1' based on the drive amplitude. $P$ is an external drive of a fixed amplitude provided to the oscillator, known as the pump. In this study, we specifically focus on the design of mechanical NAND gates, since they are universal gates that enable all Boolean operations \cite{Natarajan_20}. The schematic of this logic gate is shown in Figure~\ref{NAND diagram}a). Previous work on nanomechanical logic has generally used the jump-up transition~\cite{Yao_14, Tadokoro_21}. We therefore first focus our attention on this parameter regime.

 A NAND gate functions by producing a binary `0' output when provided with two binary `1' inputs. For all other combinations of inputs, a NAND gate produces binary `1' outputs, as shown in Table \ref{NAND Truth}. To ensure the Duffing oscillator's behaviour is analogous to a NAND gate, the input drives $A$ and $B$ must be configured such that when they are both `1', the sum of all input drives is below the jump up threshold, $\tilde{F}_{c,up}$. Then, the oscillator is expected to produce a logical `0' output, corresponding to low displacement. Similarly, when the input drives are any other combinations of binary signals, the oscillator must be driven above the jump threshold to produce a logical  `1' output, corresponding to a high displacement. This can be achieved by manipulating the phases and amplitudes of the input sinusoidal drives (i.e. $A$, $B$ and $P$). \\

To achieve the NAND gate functionality, we define the amplitude of $P$ to be above the jump threshold. In addition, we define the phase of $P$ to be $\pi$ out of phase with both $A$ and $B$. This creates destructive interference between $A$ and $P$ as well as between $B$ and $P$. The amplitudes of these drives is chosen such that the oscillator only exhibits a low drive response (`0') when both $A$ and $B$ are `1'. From these conditions, we find the input drives must satisfy the following mathematical relationships:
\begin{align}
\left|\tilde{F}_P - \tilde{F}_{\text{`}0\text{'}}-\tilde{F}_{\text{`}0\text{'}} \right|& >\tilde{F}_{c,up} \label{Gate2}\\
\left|\tilde{F}_P - \tilde{F}_{\text{`}1\text{'}} - \tilde{F}_{\text{`}0\text{'}} \right|&> \tilde{F}_{c,up} \label{Gate3}\\
\left|\tilde{F}_P - \tilde{F}_{\text{`}1\text{'}} -\tilde{F}_{\text{`}1\text{'}} \right|& <\tilde{F}_{c,up}, \label{Gate4}
\end{align}
where $\tilde{F}_P,  \tilde{F}_{\text{`}0\text{'}}$ and $\tilde{F}_{\text{`}1\text{'}}$ respectively denote the drive amplitudes of the pump, the binary `0' signal, and the binary `1' signal. In doing so, we expect the Duffing oscillator to behave like a NAND gate. For the remainder of this paper, this mechanical gate will be represented with the conventional NAND gate symbol, which is shown in Figure~\ref{NAND diagram}b). 

\begin{table}[h!]
\begin{center}
\begin{tabular}{|c|c|c|}
\hline
Input A & Input B & Output \\ \hline
0       & 0       & 1      \\ \hline
0       & 1       & 1      \\ \hline
1       & 0       & 1      \\ \hline
1       & 1       & 0      \\ \hline
\end{tabular}
\end{center}
\caption{ Truth table of a NAND logic gate. }
\label{NAND Truth}
\end{table}

\begin{figure*} [ht!]
\begin{center}
\includegraphics[width=0.75\textwidth]{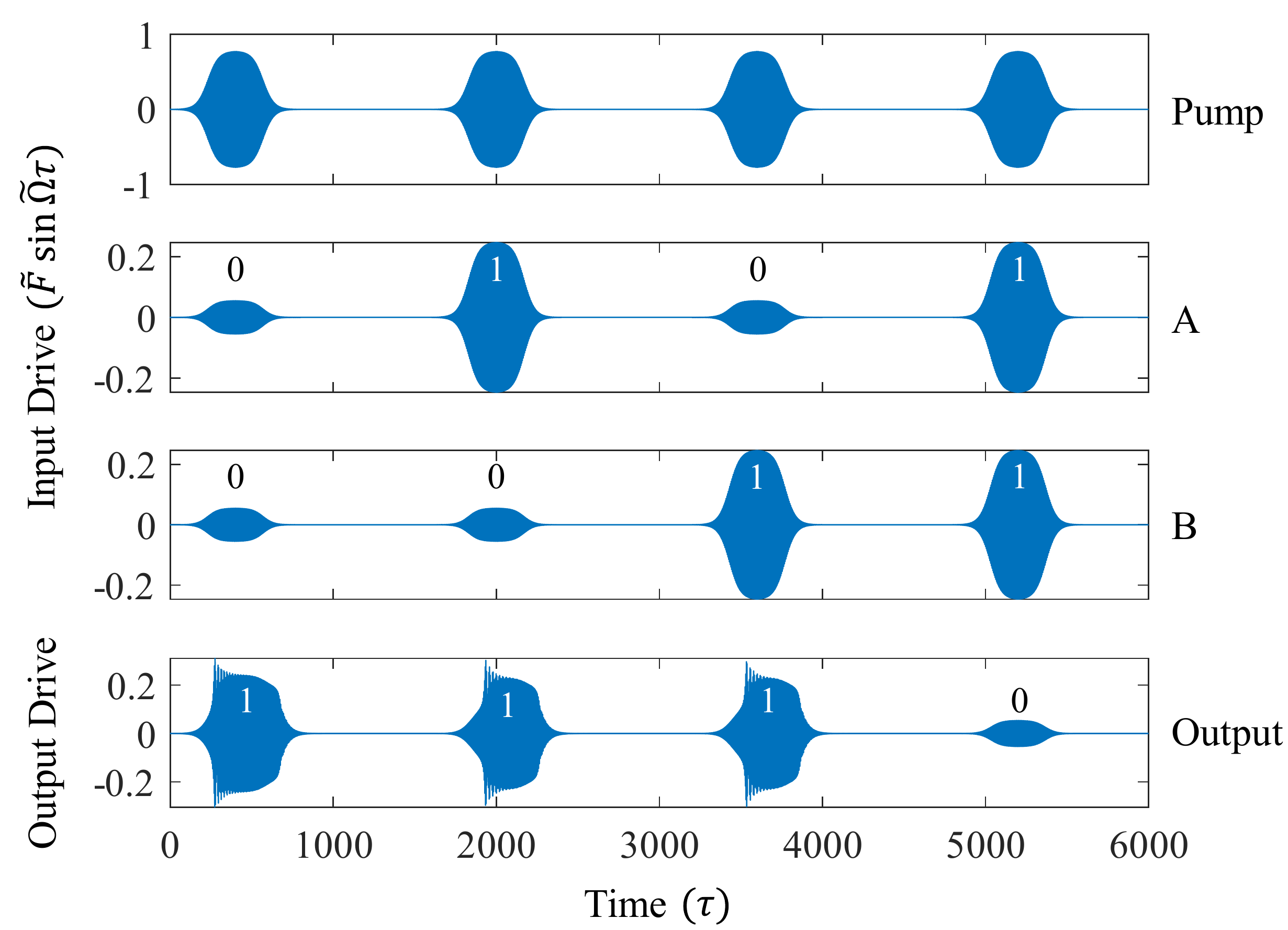} 
\end{center}
  \caption{Visual truth table of a single mechanical NAND gate. The modulated inputs and the simulated output of a single mechanical NAND gate are shown. Each pulse is labelled as a `1' or `0' binary signal. The parameters used to construct this figure were: $Q = 20, \tilde{F}_{\text{`}0\text{'}} = 0.056, \tilde{F}_{\text{`}1\text{'}} = 0.25, \tilde{F}_P = 0.78 , \tilde{F}_{c,up} = 0.42,$ $\tilde{\Omega} = 1.4$. Note: the blue pulses contain fast oscillations at the mechanical resonance frequency which cannot be resolved at these long time scales.} \label{NAND single}
\end{figure*}

Numerical simulations were developed to model the time dynamics of the mechanical NAND gate to confirm its behaviour was consistent with the truth table shown in Table \ref{NAND Truth}. To ensure the functionality of our simulated NAND gate, we carefully modulate the amplitude of the input drives such that the oscillator responds adiabatically. This is desirable as it should result in standardised input/output signal envelopes that can be shared among different logic gates within a larger circuit. To satisfy this criteria we choose a piece-wise amplitude modulation consisting of two mirrored logistic functions combined with a quadratic spline (see Appendix). We then tune the amplitudes to ensure that equations \eqref{Gate2} - \eqref{Gate4} are satisfied. The pump and input drives for a train of logical `0' and `1' states are shown in Figure~\ref{NAND single} (top three plots).
 The output force of a Duffing oscillator was then numerically solved as a function of time. This was done by using the ode45 MATLAB algorithm to solve equation \eqref{dimensionless_Duffing} given the input drives shown in Figure~\ref{NAND single} (top three plots). The result of this time dynamics simulation is shown in Figure~\ref{NAND single} (bottom plot). As shown, when both inputs are logical `1's, the oscillator produces a logical `0' output. All other combinations of inputs result in logical `1' outputs. However, we observe that the logical `1' output signals are not identical to the logical `1' input signals, despite the adiabatic design of the inputs. This is a consequence of the oscillator overshooting as it jumps up from its logical `0' state to its logical `1' state. Regardless of the overshooting, all outputs agree with the NAND gate truth table shown in Table \ref{NAND Truth}. This confirms that a single Duffing oscillator can be configured to behave as a mechanical NAND gate. Indeed, this result is consistent with previously published literature \cite{Song_19, Guerra_10}.

\section{Cascaded NAND Logic Gates}

\begin{figure*} [hbt!]
\begin{center}
\includegraphics[width=0.75\textwidth]{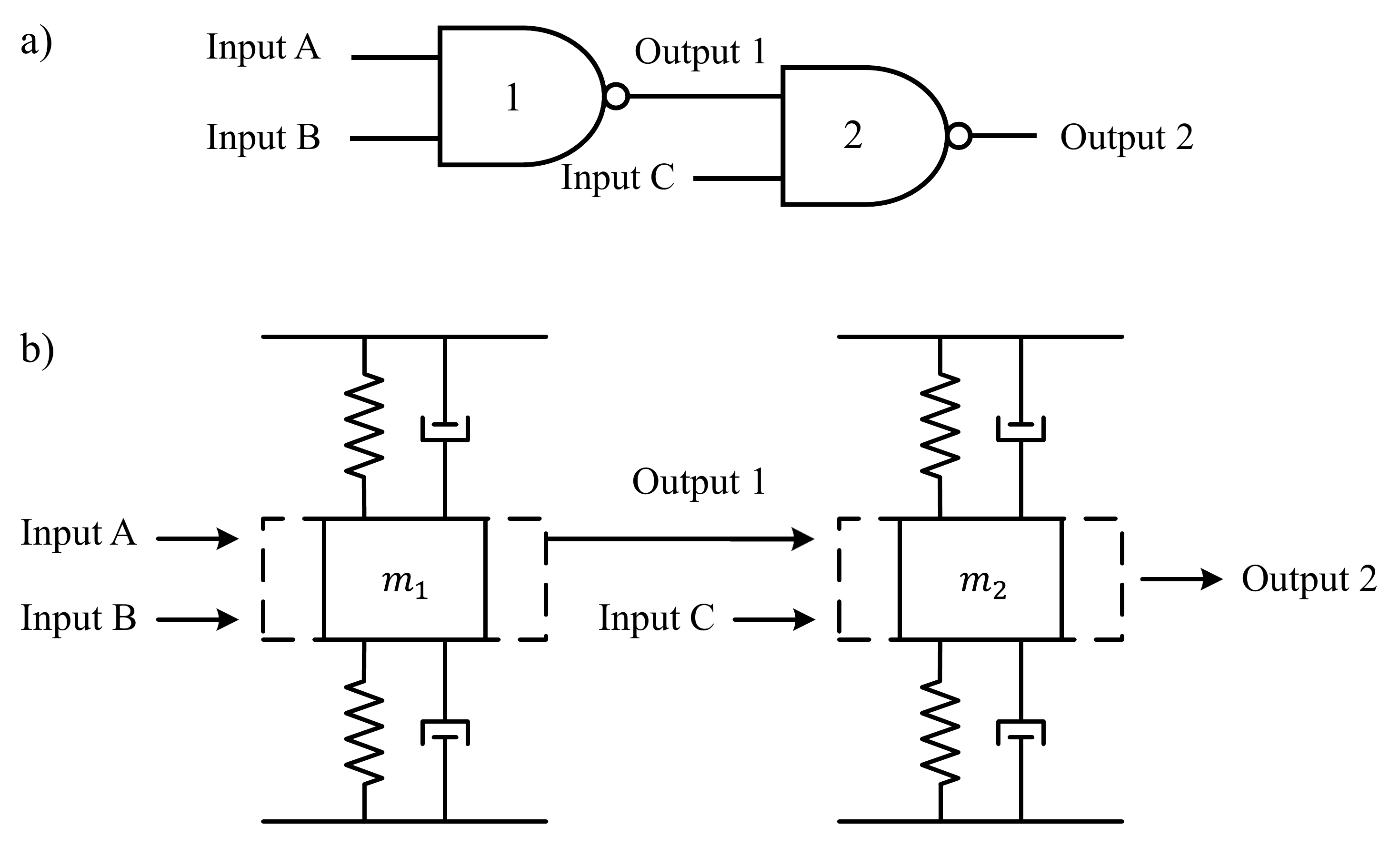} 
\end{center}
  \caption{a) Schematic of cascaded NAND gates. b) Implementation of two cascaded Duffing oscillators within numerical simulation. } \label{cascading}
\end{figure*}

In order to cascade multiple NAND gates together to form a complex circuit, the output of one gate must be used as the input for another. We consider the simplest possible scenario of two cascaded NAND gates, as shown in Figure~\ref{cascading}a). Here, the first NAND gate is provided with two inputs, $A$ and $B$, and produces one output, referred to as Output 1. Output 1 and a third input, Input C, are then used as the inputs to the second cascaded NAND gate, which produces Output 2. Larger cascaded circuits can be constructed using similar designs. Within our mechanical system, the cascading between the two oscillators is achieved by direct interference between Output 1 and Input C, as illustrated in Figure~\ref{cascading}b). That is, the oscillation of the first oscillator induces a mechanical force on the second. The final output of the cascaded gates is given by the output force of the second oscillator, as illustrated in Figure~\ref{cascading}b).
%
The equations of motion of this system is given by:
\begin{align} 
\frac{d^2 \tilde{x}_1}{d \tau^2}+ Q^{-1}\frac{d \tilde{x}_1}{d\tau} + \tilde{x}_1 +  \tilde{x}_1^3 = (\tilde{F}_A + \tilde{F}_B + \tilde{F}_P) \sin( \tilde{\Omega} \tau) \label{coup_osc1}\\
\frac{d^2 \tilde{x}_2}{d \tau^2}+ Q^{-1}\frac{d \tilde{x}_2}{d\tau} + \tilde{x}_2 +  \tilde{x}_2^3 = ( \alpha \tilde{x}_1+ \tilde{F}_C + \tilde{F}_P) \sin( \tilde{\Omega} \tau)
\end{align}
where $\alpha$ is a scaling factor that reduces the amplitude of Output $1$ to appropriate levels for the subsequent gate. 
It is important to note that we have implicitly assumed the existence of a mechanical diode between the gates that prevents backward flow of acoustic energy (i.e. there is no $\alpha \tilde{x}_2$ term in Equation~\ref{coup_osc1}). This assumption that logic flows ``downstream'' is consistent with conventional computing platforms, which often use buffer gates to isolate subsequent stages of logic. Several mechanical diodes have been proposed, with many expected to be compatible with on-chip architectures~\cite{Li_11, Popa_14, Zhang_10, Zhu_10}.
%
%
It has generally been assumed that the cascading of nanomechanical logic gates is only a technological challenge \cite{Wenzler_14,Yamaguchi_11, Guerra_10}. Therefore, we anticipated the cascading of two mechanical logic gates to be successful. 

\begin{figure*} [ht!]
\begin{center}
\includegraphics[width=1\textwidth]{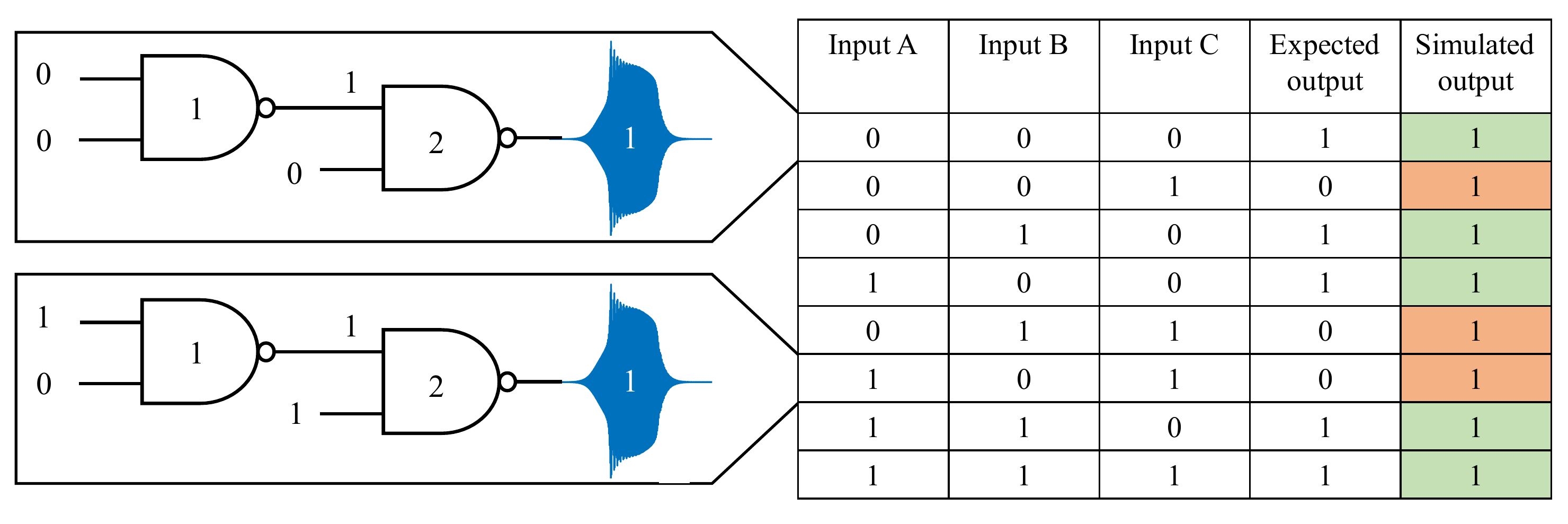} 
\end{center}
  \caption{ The simulated and expected truth table of two cascaded NAND gates using the jump up bistable transition. Green and red highlights are respectively used to indicate agreement and disagreement between the expected and simulated outputs. Two examples of the system output are shown on the left hand side of the truth table. The parameters used to construct this figure were: $Q = 20$, $\tilde{F}_{\text{`}0\text{'}} = 0.056$, $\tilde{F}_{\text{`}0\text{'}} = 0.25$, $\tilde{F}_P = 0.78$, $\tilde{F}_{c,up} = 0.42$, $ \tilde{\Omega} = 1.4$.} \label{NAND cascaded up}
\end{figure*}

The simulated and the expected truth table of two cascaded NAND gates and two example outputs are shown in Figure~\ref{NAND cascaded up}. Within Figure~\ref{NAND cascaded up}, the simulated outputs disagree with the expected outputs for 3 different combinations of inputs; those where both Output 1 and Input C are logical state `1'. This suggests, somewhat surprisingly, that although a single nonlinear oscillator can be configured to behave as a mechanical logic gate, the nonlinear dynamics prevent cascaded logic.

\subsection{The Problem}

\begin{figure}[hbt!]
\begin{center}
\includegraphics[width = 0.6 \linewidth]{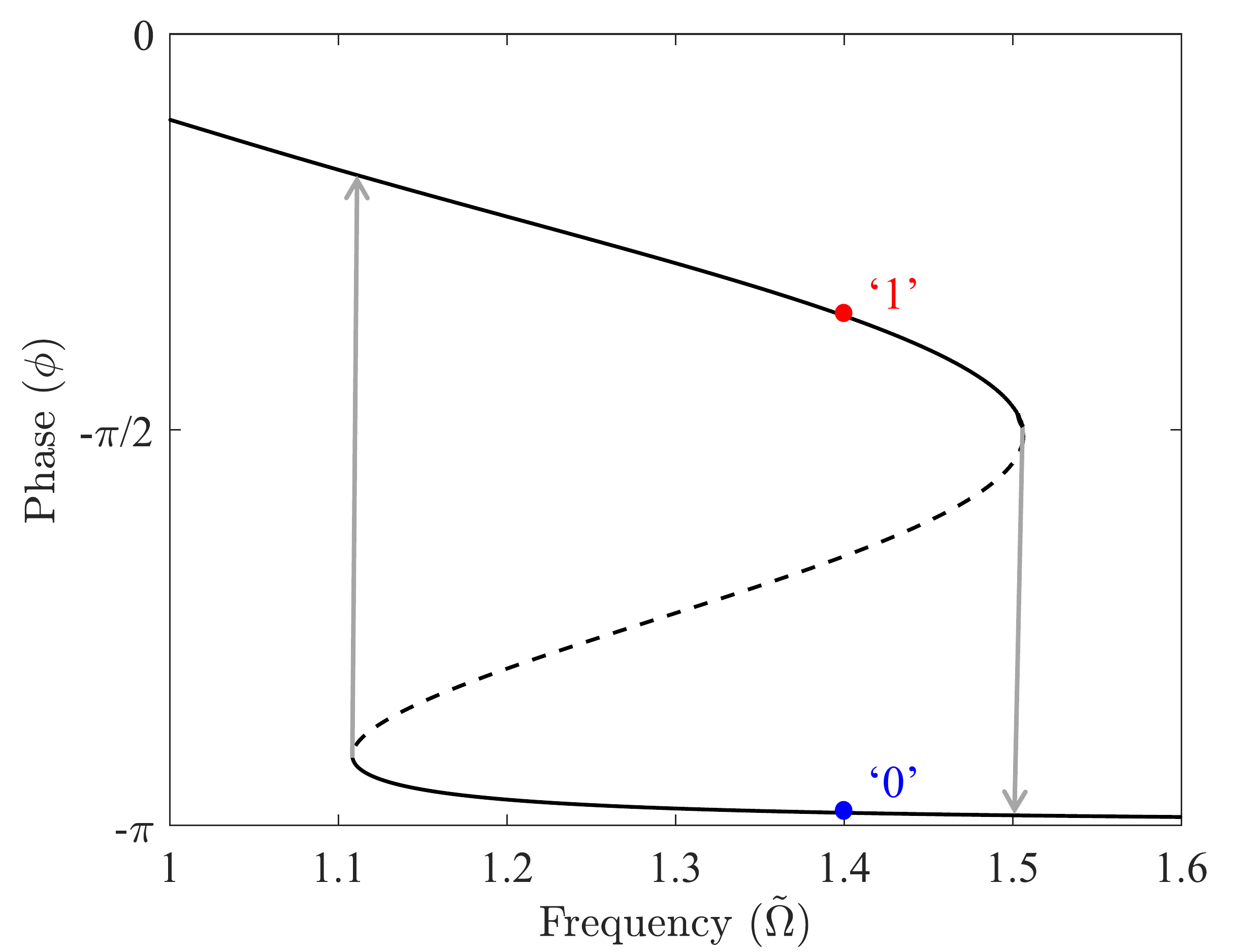}
\end{center}
\caption[Phase response curve of a Duffing oscillator (modified from \cite{Tamas_11})]{Phase response curve of a Duffing oscillator, showing the variation of phase response between the oscillation and the drive, modified from \cite{Tamas_11}. Stable solutions are plotted in solid lines, and unstable solutions in dashed lines. Up/down grey arrows are used to represent the jump up/down dynamics. 
} 
\label{phase}
\end{figure}

The inability to implement cascaded logic was found to be a consequence of the transient effect arising from the jump. The transient effect can be understood by considering the phase response of a Duffing oscillator, as shown in Figure~\ref{phase}, where $\phi$ is defined as the phase between the oscillation and the external drive. From the Figure~\ref{phase}, it can be observed that as a consequence of the bistability of the Duffing equation, the oscillator's phase response consists of two bifurcated branches of stable solutions. Within the bistable region, two stable phase relationships can be observed at a fixed frequency. In the context of our mechanical logic gates, one stable solution is used to represent a logical `0', while the other one is used to represent a logical `1', as labelled in Figure~\ref{phase}. A transition between a `0' and a `1' state requires a rephasing between the oscillation and the external drive. This re-phasing occurs over a finite time period, over which the oscillation frequency transits from one stable frequency to another. We refer to this as transient rephasing.

The functionality of a single nanomechanical NAND gate is ensured by using input drives that are perfectly in phase or out of phase with each other. To achieve a jump up, the Duffing oscillator is initialised in its `0' state with low displacement. The oscillator can either remain in the `0' state, or transition into the `1' state. When the input drives are below the jump up threshold, the oscillator will remain in its `0' state. In this case, the oscillator does not experience transient rephasing, because it does not jump. However, when the drives are above the jump up threshold, the oscillator will jump from a `0' to a `1' state, and experience transient rephasing. As a result, when performing NAND gate operations using the jump up phenomenon, binary `1' outputs are observed to have rapid phase variations. 
When attempting to cascade these outputs, the transient rephasing prevents the desired superposition between the inputs and pump, and the jump up threshold ($\tilde{F}_{c,up}$) is briefly exceeded. In essence, the short overshooting of the drive force enables the oscillator to latch to the upper branch of the bistability. This occasionally results in logical `1' outputs from cascaded system, rather than anticipated `0' outputs.

The transient effect of the jump has not been previously discussed by any other literature in the context of mechanical computing. This is because most previous literature only investigated the steady state dynamics of the Duffing oscillator \cite{Schmid,Brennan_07,Lifshitz}, such as the drive response shown in Figure~\ref{jumps}. Our simulations show that although the jump up transient effect does not affect the functionality of a single gate, it does preclude further cascading of mechanical NAND gates. This is concerning, since this potentially implies that it is impossible to cascade mechanical logic gates and hence perform complex computational tasks.   

\subsection{Solution}

While previous nanomechanical computing protocols based on Duffing oscillators have generally used the jump-up transition \cite{Tadokoro_21}, an alternative is to rely on the jump-down transition. Here, we investigate whether this can mitigate the transient effects and enable cascaded logic. In this case, the oscillator must be initialised in its `1' state to leverage the jump down phenomenon. Then, when provided with two binary `1' input drives, the oscillator should jump down to its `0' state. For all other combinations of inputs, the oscillator should remain in its `1' state, achieving the NAND gate functionality. In doing so, the transient effect should only appear on the binary `0' outputs. Since `0' signals have much smaller amplitudes, we expect the impact of the transient effect to be reduced. 

To verify the above hypothesis, we simulated the outputs of cascaded mechanical NAND gates using jump down. Previously, when utilising the jump up phenomenon, the oscillators could be easily initialised with zero position and momentum (i.e., `0' state). However, it is less trivial as to how to initialise the oscillators in their `1' states, as required to achieved jump down. One way to do this is to consistently provide the oscillators with very strong drives such that they equilibrate in their `1' states. A more optimal way to initialise the oscillators in their `1' states is to provide them with input drives that vary adiabatically in both amplitude and frequency. The amplitude and frequency modulation is discussed in further detail in the Appendix. Note that this modification does not impact the functionality of a single NAND gate.  

\begin{figure*} [ht!]
\begin{center}
\includegraphics[width=1\textwidth]{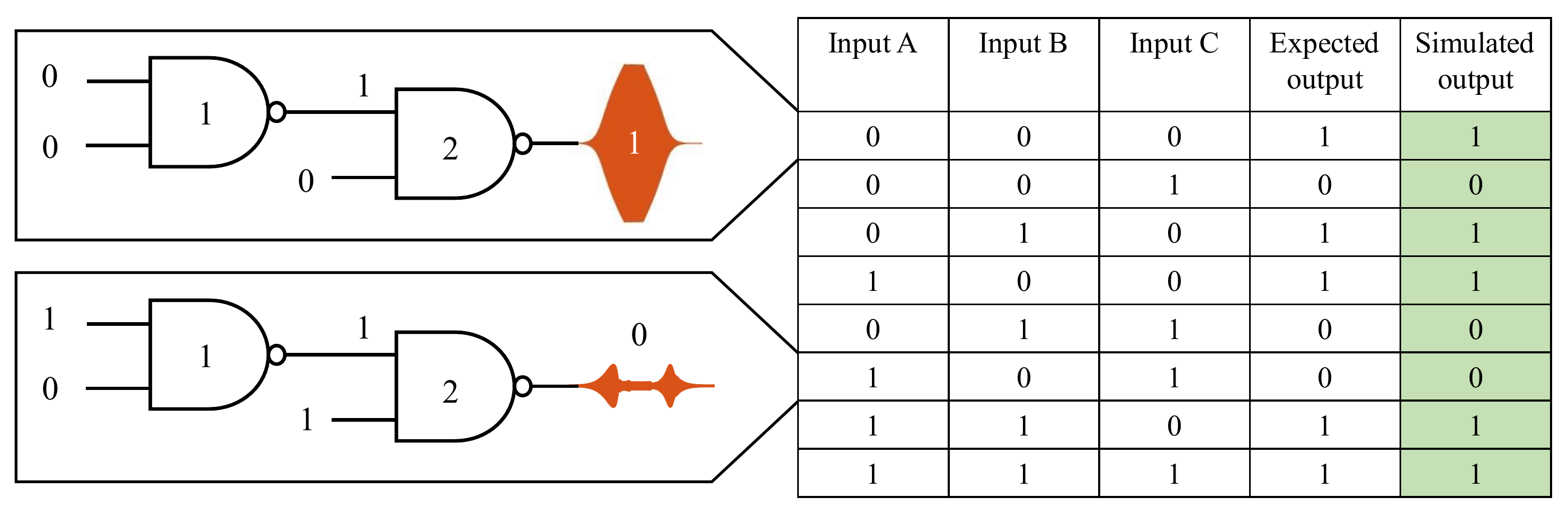} 
\end{center}
  \caption{ The simulated and expected truth table of two cascaded NAND gates using jump down. Green highlights are used to indicate agreement between the expected and simulated outputs. Two examples of the system output are shown on the left hand side of the truth table. The parameters used to construct this figure were: $Q = 10, \tilde{F}_0 = 0.0, \tilde{F}_1 = 0.17, \tilde{F}_P = 0.35 , \tilde{F}_{c,down} = 0.15, \tilde{\Omega} = 1.36$. 
  }
  \label{NAND cascaded down}
\end{figure*}

Figure~\ref{NAND cascaded down} displays the simulation truth table of cascaded NAND gates when using jump down. Similarly to Figure~\ref{NAND cascaded up}, two examples of simulated outputs are displayed, along with their expected outputs. Unlikely previously, the transient effect is only present on the binary `0' outputs, as shown in Figure~\ref{NAND cascaded down}. Furthermore, by using jump down to perform computation, the desired outputs are obtained for all possible combinations of inputs. Thus, cascaded NAND gates do function as desired when using jump down, confirming our hypothesis.

\section{Further Cascading}

\begin{figure*} [ht!]
\begin{center}
\includegraphics[width=0.91\textwidth]{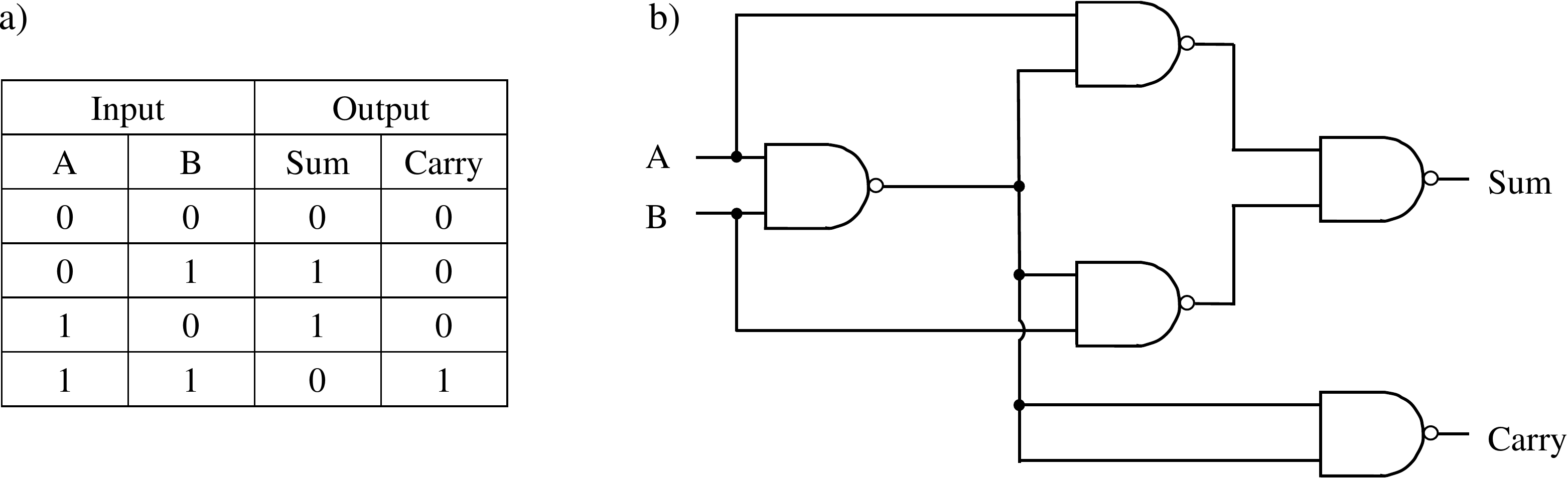} 
\end{center}
  \caption[Truth table and circuit diagram of a half-adder]{a) Truth table of a half adder, taken from \cite{Natarajan_20}. b) Circuit diagram of a half adder. }
  \label{half adder}
\end{figure*}

To test whether cascaded jump-down logic can be used for simple calculations involving a larger number of gates, we simulate a half adder. The truth table of a half adder is shown in Figure~\ref{half adder}a), and it can be constructed using a series of five cascaded NAND gates, as shown in Figure~\ref{half adder}b). As a demonstration that the cascading of NAND gates is still successful on a larger scale, we simulated the output of a mechanical half adder, using the design shown in Figure~\ref{half adder}b).

\begin{figure*} [h!]
 \centering
\includegraphics[width=0.7\textwidth]{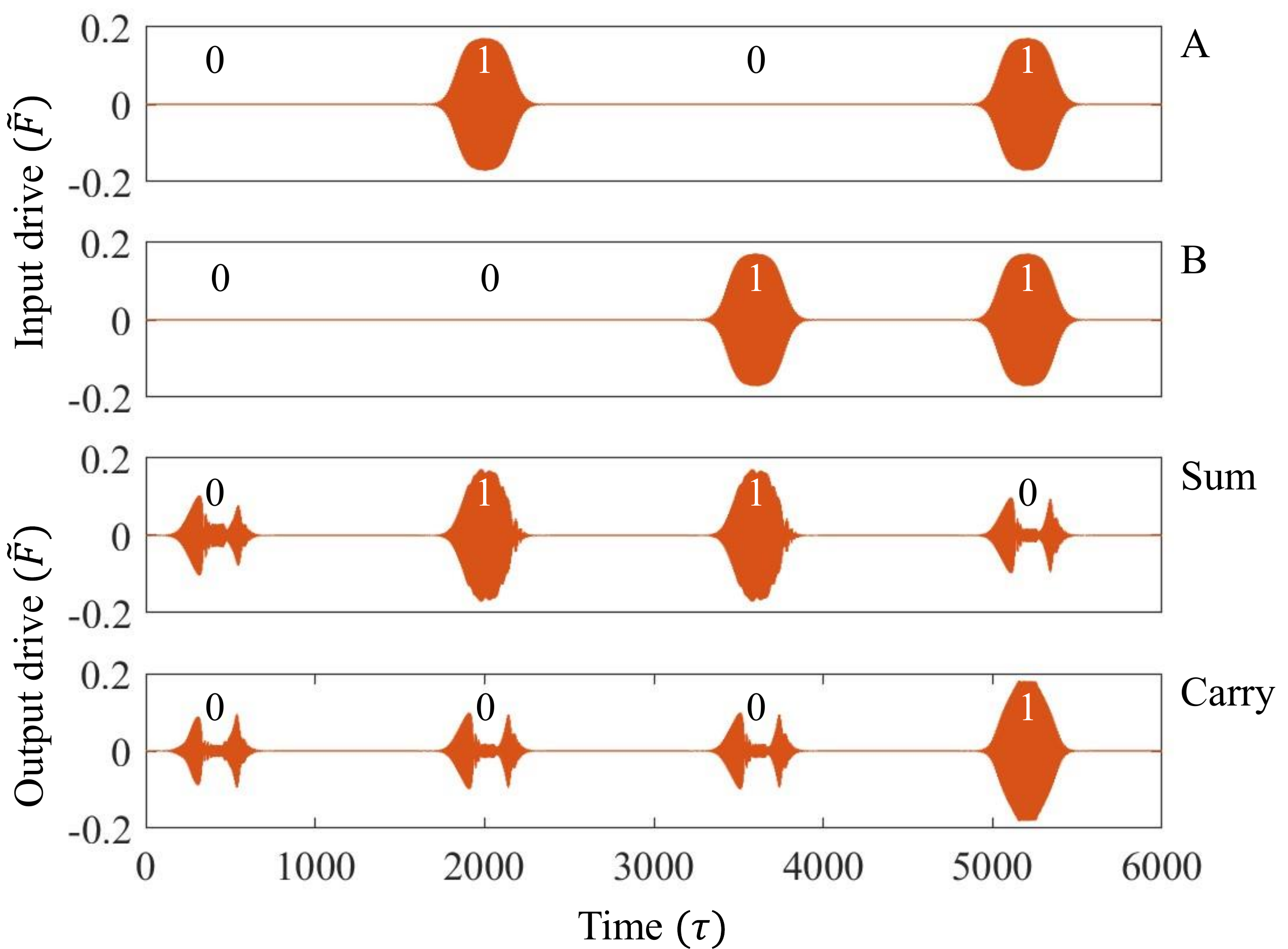} 
  \caption[Simulation results of a mechanical half adder]{ Simulation results of a mechanical half adder. The parameters used to construct this figure were: $Q = 10, \tilde{F}_0 = 0.0, \tilde{F}_1 = 0.17,$ $ \tilde{F}_P = 0.35 , \tilde{F}_{c,down} = 0.15, \tilde{\Omega} = 1.36$.}
  \label{half adder results}
\end{figure*}

Similar to previously, we used sinusoidal drives as inputs, and numerically solved for the output forces of the half adder. The result is shown in Figure~\ref{half adder results}. The four sub-figures of  Figure~\ref{half adder results} represent input A, input B, output sum, and output carry of the half adder, respectively. It is observed that the outputs of the mechanical half adder are consistent with the truth table shown in Figure~\ref{half adder}a). This means that the mechanical half adder functions as desired, confirming that our cascading of NAND gates works at larger scales.

\section{Conclusion}
In this paper, we explored the possibilities of constructing scalable mechanical logical systems using Duffing oscillators. We have identified two methods to perform mechanical computation, those being the jump up and jump down phenomena exhibited by Duffing oscillators. Using numerical simulations, we have demonstrated that by taking advantage of either phenomenon, a single oscillator can be manipulated to behave as a mechanical logic gate.

Interestingly, we have also observed that the jump phenomenon gives rise to transient effects, which have not been discussed in other literature in the context of nanomechanical computing. Although the jump up transient effect does not interfere with the functionality of a single gate, it does appear to preclude further cascading of NAND gates. To solve this problem, we have proposed to perform computation by taking advantage of the jump down phenomenon combined with adiabatic frequency and amplitude modulation. Only then, can the transient effect be minimised such that the cascading of logic gates works as desired.

\section{Acknowledgments}

This research was primarily funded by the Australian Research Council and the Lockheed Martin Corporation through the Australian Research Council Linkage Grant No. LP160101616. Support was also provided by the Australian Research Council Centre of Excellence for Engineered Quantum Systems (Grant No. CE170100009). W.P.B, G.I.H. and C.G.B. acknowledge fellowships from The University of Queensland (Grant No. UQFEL1719237) and the Australian Research Council (Grants No. DE210100848 and No. DE190100318), respectively. The authors would like to thank Luke Uribarri and Ned Allen for insightful discussions and starting us down this path.

\bibliography{ijuc}

\section*{Appendix}

\textbf{
Single gate simulation (jump up) details:}

A piece-wise logistic modulation was applied to ensure that the amplitude of the input drives vary slowly and also reach their maximum amplitude for a sustained period of time, as shown in Figure~10a). This was achieved by splicing two mirrored logistic equations together. The equation of a logistics function that centres around $\tau_c$ is given by:
\begin{align*}
A(\tau) &= \frac{\tilde{F}}{1+ e^{-\kappa (\tau-\tau_c)}}
\end{align*}where $\tilde{F}$ is the maximum amplitude of the function, and $\kappa$ controls the function growth rate.  Note that the modulation would not be differentiable at the centre if it only consisted of two mirrored logistic functions. To ensure differentiability, a third quadratic spline was added. The final amplitude modulation, $A(\tau)$, used for a pulse starting at $\tau = 0$ in our simulation was given by:
\[   A(\tau) =   \left\{
\begin{array}{ll}
      \frac{\tilde{F}}{1+ e^{-\kappa (\tau -\tau _c)}} & 0 \leq \tau \leq (\frac{1}{2} - \epsilon_A) \tau_1\\
     - \frac{(\tau - \tau_{1}/2)^2}{a}+h & (\frac{1}{2} - \epsilon_A) \tau_1 \leq \tau \leq (\frac{1}{2} + \epsilon_A) \tau_1 \\
      \frac{\tilde{F}}{1+ e^{\kappa (\tau +\tau_c - \tau_{1})}} &(\frac{1}{2} + \epsilon_A) \tau_1 \leq \tau \leq \tau_1 \\
\end{array} 
\right. \]where $\tau_{1}$ is the length of the pulse, $\epsilon$ controls the length of the quadratic spine, $a$ and $h$ respectively represent the gradient and the height of the quadratic spline. For a pulse with length of $\tau _1 = 800$, the optimised values of  $\kappa, \epsilon_A $ and $\tau_c$ were found to be $0.03$, $0.02$, and $\ln(1000)/\kappa$, respectively. The values of $a$ and $h$ were chosen to ensure continuity and differentiability at the logistic endpoints. In our simulation, the exact values used were:
\begin{align*}
a = \frac{16 (1+1000 \exp(-11.52))^2}{15 \tilde{F} \exp(-11.52)}\\
 h = \frac{1+1240 \exp(-11.52)}{(1+1000 \exp(-11.52))^2} \tilde{F}
\end{align*}
Note that the value of $\tilde{F}$ depends on the type of input signals (`0', `1' or pump). Finally, the modulated input pulses are given by $A (\tau) \sin (\tilde{\Omega} \tau)$, where $\tilde{\Omega}$ is a fixed frequency. \\

\textbf{Cascaded gate simulation details}:

To simulate multiple cascaded NAND gates, the same amplitude modulation as shown above was applied to new inputs. When taking advantage of the jump up phenomenon, no frequency modulation of the input pulses was required. However, to achieve the jump down phenomenon, a piece-wise frequency modulation function, $f(\tau)$, was also applied to input pulses. That is, the input pulses are instead given by $A(\tau) \sin (f(\tau) )$, and the frequency of the input pulses is given by $f'(\tau)$. A desirable frequency modulation should satisfy two criteria; the frequency of the input pulses should remain fixed at $\tilde{\Omega}_\mathrm{max}$ for a sustained period of time, and ramp to and from $\tilde{\Omega}_\mathrm{max}$ steadily, see Figure~10b). This implies that the modulation must be divided into three parts; increasing frequency ($0 \to \tilde{\Omega}_\mathrm{max}$), fixed frequency ($\tilde{\Omega}_\mathrm{max}$) and decreasing frequency ($\tilde{\Omega}_\mathrm{max} \to 0)$. An obvious choice of center part of the modulation is $f( \tau) = \tilde{\Omega}_\mathrm{max} \tau$. For the first part of the piece-wise modulation, $f (\tau)$ must give rise to continuity and differentiability at both its left and right hand end points. To satisfy these four conditions, the simplest form $f(\tau)$ can take in this region is a cubic function. By the same rationale, $f(\tau)$ must also be a cubic function in the third part of the modulation. Therefore, the final form of $f(\tau)$ is given by:

\[   f(\tau) =   \left\{
\begin{array}{ll}
     a_1 \tau ^3 + b_1 \tau^2 + c_1 \tau + d_1 & 0 \leq \tau \leq  (\frac{1}{2} - \epsilon_f) \tau_1 \\
   \tilde{\Omega}_\mathrm{max} \tau & (\frac{1}{2} - \epsilon_f) \tau_1 \leq \tau \leq (\frac{1}{2} + \epsilon_f) \tau_1 \\
 a_2 \delta\tau ^3 + b_2 \delta\tau^2   + c_2  \delta\tau + d_2 &(\frac{1}{2} + \epsilon_f) \tau_1 \leq \tau \leq \tau_1 \\
\end{array} 
\right. \]
where $\delta \tau=\tau_1-\tau$. Within our simulation, the values of each parameter were chosen to be:
\begin{align*}
\epsilon_f &= \frac{1}{16}, \tau_1 = 800, \tilde{\Omega}_\mathrm{max} = 1.36\\
a_1 &= -\frac{17}{1531250} +\frac{19 \pi}{5359375}, b_1 = \frac{34}{4375} -\frac{57 \pi}{30625}, c_1 = 0, d_1 = 76 \pi\\
a_2 &= - \frac{17}{428750} + \frac{27 \pi}{2143750}, b_2 = \frac{578}{30625} - \frac{81 \pi}{12250}, c_2 = 0, d_2 = 270\pi
\end{align*}

Note that this is not the only combination of parameters that ensure continuity and differentiability. The above set of values were chosen such that $a_1$ and $a_2$ are absolutely minimised, and the frequency of the input pulse (given by $f'(\tau)$) varies as linearly as possible, as shown in Figure~10b). This satisfies the latter of earlier mentioned key criteria. 
 \begin{figure*} [ht]
\centering
\includegraphics[width=1\textwidth]{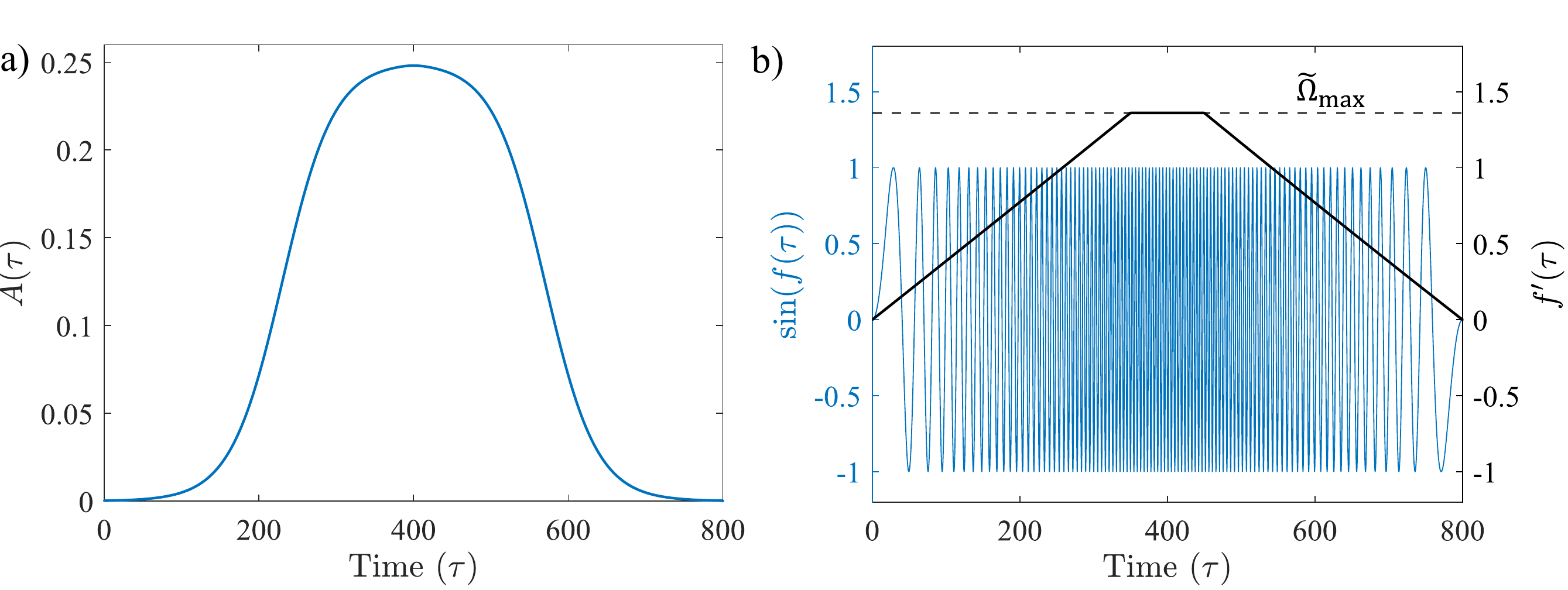} 
  \caption{Piece-wise modulation of input pulses a) Amplitude modulation used for all inputs (both jump up and jump down) b) Frequency modulation used to achieve the jump down phenomenon.}
  \label{modulation}
\end{figure*}

\end{document}